\documentclass[twoside,a4paper]{article}

\usepackage{pstricks}
\usepackage{graphicx}
\usepackage{epsf,psfig,multicol,pst-grad,color}

\def\5{\hat }

\newcommand{\be}{\begin{equation}}
\newcommand{\ee}{\end{equation}}

\newcommand{\bea}{\begin{eqnarray}}
\newcommand{\eea}{\end{eqnarray}}

\newcommand{\nn}{\nonumber\\ }

\newcommand{\0}{\over }
\newcommand{\2}{{1\over2}}

\newcommand{\g}{g_{\rm eff}}
\newcommand{\geff}{g_{\rm eff}}

\newcommand{\muMS}{\bar\mu_{\rm MS}}

\newcommand{\re}{{\rm Re}}

\newcommand {\Tr}{\mathrm{Tr}}

\def\cf{C_{\rm F}}

\def\Tr{{\,\rm Tr\,}}
\def\Im{{\,\rm Im\,}}
\def\Re{{\,\rm Re\,}}
\def\tr{{\,\rm tr\,}}

\def\slashchar#1{\setbox0=\hbox{$#1$}           
   \dimen0=\wd0                                 
   \setbox1=\hbox{/} \dimen1=\wd1               
   \ifdim\dimen0>\dimen1                        
      \rlap{\hbox to \dimen0{\hfil/\hfil}}      
      #1                                        
   \else                                        
      \rlap{\hbox to \dimen1{\hfil$#1$\hfil}}   
      /                                         
   \fi}

\usepackage{bm}

\def\intK{\int_K}

\def\sumint{\hbox{$\sum$}  \!\!\!\!\!\!\!\int}

\def\G{G}

\definecolor{MyViolet}{rgb}{1.0,.0,1.0}
\definecolor{MyGreen}{rgb}{0.,0.7,0.}

\title{The entropy of hot QCD at large $N_f$: \\ Successfully testing weak coupling techniques}
\author{Jean-Paul Blaizot and Andreas Ipp \\
\small{ECT*, Villa Tambosi, Strada delle Tabarelle 286,}\\
\small{I-38050 Villazzano Trento, Italy }\\
\\
Anton Rebhan and Urko Reinosa \\
\small{Institut f\"ur Theoretische Physik, Technische Universit\"at Wien, }\\
\small{Wiedner Hauptstr.~8-10, A-1040 Vienna, Austria}}
\date{}
\begin {document}
\maketitle
\begin{abstract}
It has been known for some time that the entropy of hot QCD 
is well reproduced by weak coupling techniques. These do not 
identify with perturbation theory, known to have poor convergence 
properties, but involve resumming the physics of hard thermal loops in
 the non-perturbative $\Phi$-derivable two-loop approximation.
 We test this approximation scheme in the limit of large flavor number ($N_f$) where the exact result can be calculated for a wide range of couplings
 for which the influence of the Landau pole is negligible.
 Using full momentum-dependent fermionic self-energies instead of the previously
 only known weighted average value to order $g^3$, the exact result for the entropy can
 be remarkably well reproduced by the HTL resummed theory for a natural
 choice of the renormalization scale, and this up to large values of the
 coupling. This gives confidence in the reliability of weak coupling techniques when
 applied to the thermodynamics of QCD, at least for temperatures $T\geq3T_{c}$.

\end{abstract}

\section*{Introduction}

The entropy of hot QCD 
is well reproduced 
for temperatures $T\geq3T_{c}$
by the non-perturbative $\Phi$-derivable two-loop approximation \protect\cite{BIR}.
Its success has been established by comparison to lattice simulations \cite{Karsch}.
Here we want to present an independent test \cite{Blaizot:2005wr} of 
this approximation scheme  
in the limit of large flavor number $N_f$
\cite{Largenf}.

Approximations based on the $\Phi$ functional 
are constructed from the two-particle-irreducible (2PI) skeleton expansion
\cite{Baym:1962}
where the thermodynamic potential is expressed in
terms of dressed propagators 
according to
\bea\label{LWQCD}
\Omega[\G,S]/T&=&\2 \Tr \log \G^{-1}-\2 \Tr \Pi \G\nn
&&
- \Tr \log S^{-1} + \Tr \Sigma S + \Phi[\G,S] \, ,
\eea
where ``Tr'' refers to full functional traces 
and
$\Phi[\G,S]$ is the sum of 
2PI ``skeleton''
diagrams.
The self-energies $\Pi[G,S]=\G^{-1}-\G^{-1}_0$ and 
$\Sigma[G,S]=S^{-1}-S^{-1}_0$
are determined by the stationarity property 
${\delta \Omega[\G,S]/\delta \G}=0={\delta \Omega[\G,S]/\delta S}$.
For the entropy ${\mathcal S}=(\6P/\6T)_{\mu}$
with $P=-\Omega/V$
one obtains 
\cite{BIR}:
			\bea
			\label{S2loop}
			{\mathcal S}&=&-\tr \intK{\6n(k_0)\0\6T} \left[ \Im 
			\log \G^{-1}-\Im \Pi \Re \G \right] \nn
			&&-2\tr \intK{\6f(k_0)\0\6T} \left[ \Im
			\log  S^{-1}-\Im\Sigma \Re S \right].
			\eea

\begin{figure}
\begin{center}\includegraphics{
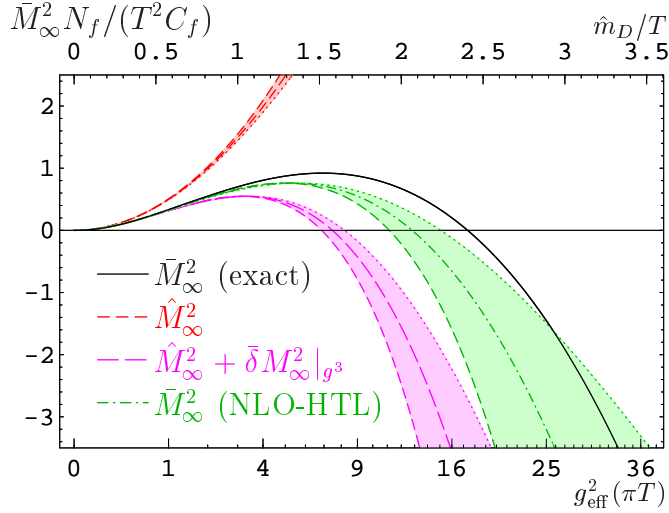}\end{center}
\caption{Comparison of the averaged asymptotic thermal quark mass squared,
$\bar M_{\infty}^{2}$, in various approximations. 
In this and
the following plots, the renormalization
scale $\muMS$ is varied around the FAC-m scale by factors of 2. 
\label{fig:minfinity}}
\end{figure}

In the large $N_f$ limit for QCD \cite{Largenf}, 
the effective coupling 
$\geff^2 = g^2N_f/2$ 
is
kept finite as the limit $N_f\to\infty$
is taken.
The thermodynamic potential to order $N_f^0$ and to all orders in $\geff$
is then given by a set of ring-diagrams which can be resummed.
Large-$N_f$ QCD is no longer asymptotically free.
Its renormalization scale dependence is determined
(non-perturbatively) by the one-loop beta function 
$\beta(\geff^2)=\geff^4/(6\pi^2)$
which implies a Landau singularity.
The two-loop entropy formula (\ref{S2loop}) reduces in the large $N_f$ limit to
\bea
\label{Slnf}
{\mathcal S}-{\mathcal S}_0&=&-\tr \intK{\6n(k_0)\0\6T} \left[ \Im 
\log \G^{-1}/\G_0^{-1}-\Im \Pi \Re \G \right] \nn
&&-2\tr \intK{\6f(k_0)\0\6T} \left[ 
\Re \Sigma \Im S_0 \right].
\eea
Using the 1-loop self-energy $\Pi$, this formula 
actually reproduces the large $N_f$ entropy to all orders in $\geff$.
The fermionic self-energy $\Sigma$ from this formula only has to be evaluated on the light-cone. The bare self-energy $\Sigma_{\rm b}$ is given by
\begin{equation}\label{eq:fermion}
\Sigma_{\rm b}(K)=\Sigma_{\rm th}+\Sigma_{\rm b,\,vac}=-g^2\cf\sumint_{Q} \; \, \gamma^\mu\,S_0(Q+K)\,\gamma^\nu\,\G_{\mu\nu}(Q)\,.
\end{equation}
The thermal piece $\Sigma_{\rm th}$ of the $q_0$ integration is UV finite, 
while the real part of the vacuum piece $\re\,\Sigma_{\rm b,\,vac}$ is in general logarithmically divergent, but finite on the light-cone. Because of the Landau pole $\Lambda_L$, we have to introduce a Euclidean invariant cutoff $\Lambda^2 < \Lambda_{L}^2$, and the analytic continuation of $\Sigma_{\rm b,\,vac}$ 
to the light-cone
requires a deformation of the integration path \cite{Blaizot:2005wr}.
The numerically demanding calculations have been performed on the ECT* Teraflop cluster.

\section*{HTL approximation of the self-consistent entropy}

The HTL effective action \cite{Braaten:1992gm}
is an effective action for soft modes with energy scales
$\sim gT$.
Inserting HTL into the entropy formula (\ref{S2loop}) gives the
correct $g^2$ contribution, but only a fraction of the plasmon term $\sim g^3$.
The larger part of the plasmon term arises from {\em hard} momentum
scales where HTL is no longer accurate, namely from corrections to the leading-order asymptotic masses
\be\label{minfty}
\5\Pi_{\rm T}(k_0=k)=\5 m_\infty^2=\2 \5m_D^2\, ,\qquad
\5\Sigma_\pm(k_0=\pm k)={\5 M_\infty^2\02k}
={\5 M^2\0 k}\, .
\ee
The NLO corrections to the asymptotic thermal masses are
nontrivial functions of the momentum,
\be\label{dPiSigma}
\delta m_\infty^2(k) \equiv \Re \delta\Pi_{\rm T}(k_0=k),\qquad
\delta M_\infty^2(k) \equiv \Re 
2k\, \delta\Sigma_+(k_0=k),
\ee
and involve only a single HTL propagator and no HTL vertices for hard external momentum.
These expressions can
be evaluated only numerically (NLO-HTL). However,
the weighted averages (NLA) are determined to order $g^3$
\cite{BIR}.

\begin{figure}
\begin{center}\includegraphics{
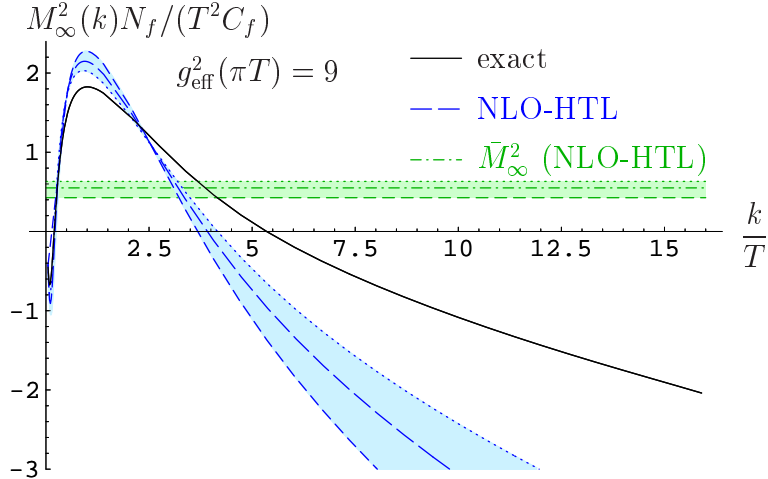}\end{center}
\caption{Asymptotic thermal quark mass squared
as a function of $k/T$ for $\geff^{2}(\pi T)=9$.
The exact large-$N_f$ 
result is compared to the NLO-HTL calculation and its relevant
average value. 
\label{fig:sigma9}}
\end{figure}

\begin{figure}
\begin{center}\includegraphics{
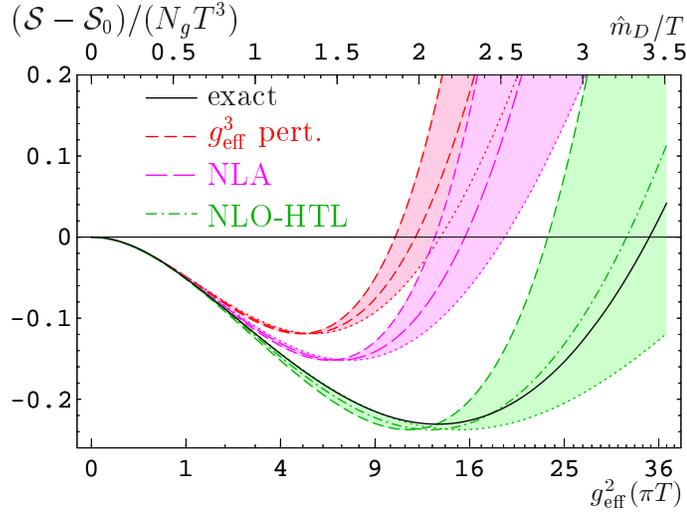}\end{center}
\caption{Entropy in the large-$N_{f}$ limit, comparing the exact large-$N_{f}$
result to the strictly perturbative expansion through order $\geff^{3}$,
the NLA result, and the NLO-HTL result. 
\label{fig:entropy}}
\end{figure}

As one can see in Fig.\ \ref{fig:minfinity}, the 
NLO-HTL-resummed result represents a considerable improvement
over the HTL-resummed result truncated at order $\g^3$
for $\g^2\geq 4$, which corresponds to $\hat m_D/T \geq 1$.
Figure \ref{fig:sigma9} shows 
the results of a numerical calculation of the asymptotic thermal quark
mass squared $M_\infty^2(k)\equiv 2k\Re\Sigma(k_0=k)$ 
for $\geff^2(\muMS\!=\!\pi T) = 9$,
normalized by $T^2C_f/N_f$. 
The exact (nonperturbative) result obtained in the large-$N_f$ limit
is given by the full lines.
In Fig.~\ref{fig:entropy} the exact and the NLO-HTL results for the entropy are
compared to simpler approximations.
It is remarkable that
a full HTL resummation using momentum-dependent asymptotic 
quark masses
can extend
the validity of the approximation in the large $N_f$ limit up to
$\hat m_D/T \sim 2.5$.
For the entropy of (small $N_f$) QCD we therefore expect likewise improvement once full HTL-resummed momentum-dependent asymptotic gluon masses have been implemented.

\end{document}